\documentclass{emulateapj}
\RequirePackage{natbib}

\def\agt{\mathrel{\raise.3ex\hbox{$>$}\mkern-14mu\lower0.6ex\hbox{$\sim$}}}
\def\alt{\mathrel{\raise.3ex\hbox{$<$}\mkern-14mu\lower0.6ex\hbox{$\sim$}}}

\newcommand{\beq}{\begin{equation}}
\newcommand{\eeq}{\end{equation}}
\newcommand{\beqn}{\begin{eqnarray}}
\newcommand{\eeqn}{\end{eqnarray}}

\shorttitle{Supermassive Black Holes}
\shortauthors{Shapiro}

\begin{document}

\title{Spin, Accretion and the Cosmological Growth of
Supermassive Black Holes} 

\author{Stuart L. Shapiro \altaffilmark{1,2}}

\affil
{\altaffilmark{1} 
Department of Physics, University of Illinois at Urbana-Champaign,
\break
Urbana, IL 61801-3080}
\affil
{\altaffilmark{2} 
Department of Astronomy and NCSA, University of Illinois at
Urbana-Champaign,
\break
Urbana, IL 61801-3080}

\begin{abstract}
If supermassive black holes (SMBHs) are the energy sources
that power quasars and active galactic nuclei, then
QSO SDSS 1148+5251, the quasar with the highest redshift ($z_{\rm QSO}=6.43$),
hosts a supermassive black hole formed within
$\sim 0.9$ Gyr after the Big Bang.  This requirement places 
constraints on the cosmological formation of SMBHs,
believed to grow from smaller initial seeds by a 
combination of accretion and mergers. We focus on gas accretion onto
seeds produced by the collapse of Pop III stars at high redshift.
We incorporate the results of recent relativistic, MHD simulations of 
disk accretion onto Kerr black holes to track the coupled evolution 
of the masses and spins of the holes.  We allow for an additional
amplification of $\sim 10^4$ in the mass of a typical seed due to 
mergers, consistent with recent Monte Carlo simulations of 
hierarchical mergers of cold, dark matter halos containing black hole seeds.
We find that the growth of Pop III black hole remnants 
to $\sim 10^9 M_{\odot}$ by $z_{\rm QSO} \ga 6.43$ favors MHD accretion disks 
over standard thin disks. MHD disks tend to 
drive the holes to a submaximal equilibrium spin rate
$a/M \sim 0.95$ and radiation efficiency $\epsilon_M \sim 0.2$,
while standard thin disks drive them to maximal
spin ($a/M = 1$) and efficiency ($\epsilon_M = 0.42$). This small difference
in efficiency results in a huge difference in  
mass amplification by accretion at the Eddington limit. The MHD 
equilibrium efficiency is consistent with the observed ratio 
of the QSO plus AGN luminosity density to the local SMBH mass density.
Our prototype analysis is designed to stimulate the 
incorporation of results from 
relativistic stellar collapse and MHD accretion simulations
in future Monte Carlo simulations of hierarchical structure formation
to better determine the cosmological role of SMBHs and 
their mass and spin distributions.

\end{abstract}




\section{Introduction}

There is substantial evidence that supermassive black holes (SMBHs)
with masses in the range  $ 10^6 - 10^{10} M_{\odot}$ exist and are the engines that
power active galactic nuclei (AGNs) and quasars 
(Rees 1984, 1998, 2001; Macchetto 1999).
There is also ample evidence that SMBHs reside at the centers
of many, and perhaps most, galaxies (Richstone et al. 1998; Ho 1999),
including the Milky Way (Genzel et al. 1997; Ghez et al. 2000, 2003;
Sch\"odel et al. 2002).
                                                                                
The highest redshift of a quasar discovered to date is $z_{\rm QSO} = 6.43$, corresponding to 
QSO SDSS 1148+5251 (Fan et al. 2003). Accordingly, if they are the energy sources in
quasars (QSOs), the first supermassive black holes
must have formed prior 
to $z_{\rm QSO} = 6.43$, or within $t = 0.87$ Gyr after the Big Bang in the concordance
$\Lambda$CDM cosmological model. This requirement sets a significant constraint on black
hole seed formation and growth mechanisms in the early universe. 
Once formed, black holes grow by a combination of
mergers and gas accretion. An important clue to the growth process is provided by the
ratio $R$ of the observed QSO plus AGN luminosity density to the 
local SMBH mass density, since  $R$ is related to the mean radiative efficiency 
$\epsilon_M$ of accretion onto black holes (Soltan 1982): 
$\epsilon_M \geq R$. Recent measurements
suggest $0.1 \la R \la 0.2$ (Yu \& Tremaine 2002; Elvis, Risaliti \& Zamorani 2002), 
a range consistent with disk accretion onto black holes.  Another clue 
is provided the estimated ratios 
$\epsilon_L$ of the 
bolometric-to-Eddington luminosities of the
broad-line quasars in
a Sloan Digital Sky Survey sample of 12,698 quasars in the redshift
interval $0.1 \leq z \leq 2.1$. This survey supports the value $\epsilon_L \approx 1$
as a physical upper limit (McLure \& Dunlop 2004). 
Barring an extraordinary coincidence, 
the range of values inferred from observations for both $\epsilon_M$ and $\epsilon_L$
suggest that a significant fraction of the mass of SMBHs
is acquired by gas accretion.  Together, these two parameters 
control the rate of growth of black holes by accretion and are crucial in 
determining whether or not 
initial seed black holes formed at high redshift have sufficient time to 
grow to SMBHs to explain quasars and AGNs.

The more massive the initial seed, the less time is required for it to grow to SMBH scale and
the easier it is to have a SMBH in place by $z \geq 6.43$. One possible progenitor
that readily produces a SMBH is a supermassive 
star (SMS) with $M \gg 10^3 M_{\odot}$ (see, e.g., Shapiro 2004 for
a recent review and references).  SMSs can form when gaseous structures 
build up sufficient radiation pressure to inhibit fragmentation and prevent normal
star formation; plausible cosmological scenarios have been proposed that can lead to
this occurence (Gnedin 2001; Bromm \& Loeb 2003).  
SMSs supported by radiation pressure will evolve in a quasi-stationary
manner to the point of onset of dynamical collapse due to a general relativistic
radial instability (Chandrasekhar 1964a,b; Feynman, unpublished, as quoted in Fowler 1964). 
The collapse of a nonrotating, marginally unstable, spherical SMS of mass $M$ 
leads directly to the formation of
a nonrotating Schwarzschild black hole of the same total mass (Shapiro \& Teukolsky 1979). 
But like most stars formed in nature, SMSs will be rotating.
In fact, in the event that viscosity and/or turbulent magnetic fields are present to 
drive these stars to uniform rotation, they are likely to be maximally 
rotating (i.e. at the mass-shedding limit) 
by the time they reach the onset of collapse (Baumgarte \& Shapiro 1999).
Recent relativistic hydrodynamic simulations have shown that unstable, maximally 
rotating SMSs of arbitrary mass $M$ inevitably collapse to SMBHs of mass $ \sim 0.9 M$
and spin parameter $a/M \sim 0.75$; the rest of the mass goes into an ambient disk
about the hole (Shibata \& Shapiro 2002: Shapiro \& Shibata 2002; Shapiro 2004).

But the fact remains that SMSs have yet to be observed and there is no concrete 
evidence that they actually form in the early universe. Moreover, 
simulations of cosmological structure
formation performed to date 
indicate that the first generation of stars are more likely to be zero-metallicity
Pop III stars in the range $10^2 -10^3 M_{\odot}$ (Bromm, Coppi \& Larson 1999, 2002; 
Abel, Bryan, \& Norman 2000, 2002; but note that Norman 2004 reports preliminary 
indications that supermassive stellar objects with $M \ga 10^4 M_{\odot}$ may be forming
as {\it second} generation stars at $10 \la z \la 15$ in his latest simulations).
So the most conservative hypothesis is that the seed black holes that later grow to become
SMBHs originate from the collapse of Pop III stars (Madau \& Rees 2001) and not SMSs, and this is the
ansatz we shall explore here. Newtonian simulations suggest that 
Pop III stars with masses in the range
$M \sim 60 - 140 M_{\odot}$ and $M \ga 260 M_{\odot}$ 
collapse directly to black holes, while stars with $M \sim 140 - 260 M_{\odot}$ undergo
explosive annihilation via pair-creation processes (Heger et al. 2003). The upper limit to the
mass of a Pop III star is set by accretion over a 
stellar lifetime, yielding $M \la 600 M_{\odot}$
(Onukai \& Palla 2003; Yoshida et al. 2003). Abel et al. (2002) argue that first generation 
stars significantly 
larger than $100 M_{\odot}$
are likely to explode before they have time to accrete to larger masses.
The Pistol Star is an example of an existing star believed to have a mass $\ga 200 M_{\odot}$, although it has high
metallicity (Figer et al. 1998).  Very massive stars are dominated 
by thermal radiation pressure, so the catastrophic collapse of those that 
do not explode will be hydrodynamically similar to
the collapse of SMSs, producing black holes with masses comparable to those of their progenitors.
The higher the mass of the black hole seed, and the earlier it forms in the universe, the easier  
it is for it to grow to a SMBH, hence the more conservative will be any
contraints imposed on the cosmological black hole growth rate
by the existence of a SMBH by $z_{\rm QSO} = 6.43$.  
Accordingly, we take the highest range of plausible
values for the masses of black hole seeds to establish the most conservative (robust) constraints,
adopting $100 \leq M/M_{\odot} \leq 600$ for the range of black hole seeds arising from
the collapse of Pop III stars at $z \leq 40$. 

The most detailed studies of SMBH formation to date involve detailed,  
Monte Carlo simulations that follow the cosmological 
growth of a distribution of black hole seeds by a combination of discrete, stochastic 
mergers as well as gas accretion (see, e.g., 
Haehnelt \& Kauffmann 2000; Volonteri, Haardt \& Madau 2003; Bromley et al. 2004; 
Haehnelt 2004; Haiman 2004;
Yoo \& Miraldi-Escud\'{e} 2004; and references
therein). Both processes are assumed to 
take place in the context of the cold dark matter (CDM) model, 
where dark matter halos merge hierarchically, and black holes are assumed to 
settle, merge and
accrete in their gaseous centers. Typically, the stellar dynamical processes 
that lead to mergers,
as well as the hydrodynamical processes that fuel accretion, are modeled in these analyses 
by implementing simple, physically plausible, rules rather than by detailed integrations of  
the governing dynamical and hydrodynamical equations of motion. Performing such  ``first principles''
integrations would prove prohibitive in this context.

In this paper we focus on SMBH cosmological growth by accretion.
We identify and explore the main variables that govern this process and 
ultimately influence 
the outcome of detailed Monte Carlo simulations that track
SMBH growth. We incorporate 
some of the most recent findings of
relativistic magnetohydrodynamical (MHD)
simulations of gas accretion onto black holes and
explore their cosmological implications for SMBH evolution.
We show how the evolution and amplification of black hole {\it mass} by accretion
is intimately tied to the evolution of black hole {\it spin}, probing some of the implications of our
earlier survey and analysis of black hole spin evolution (Gammie, Shapiro \& McKinney 2004) for  
cosmology.  We demonstrate how in principle 
the very existence of a quasar at redshift $z_{\rm QS0} = 6.43$ can help constrain 
the formation epoch and/or size of black hole seeds and select among competing 
models of accretion. 

Specifically, recent relativistic MHD simulations predict the radiation
efficiency $\epsilon_M$ as a function
of the BH spin parameter $a/M$; they also predict the spin-up rate as a
function of $a/M$.
Monte Carlo simulations that determine cosmological SMBH growth must integrate
coupled mass and spin evolution prescriptions versus time
for each accreting BH to produce reliable growth histories and final SMBH masses
and spins.  This paper provides prototype integrations of the coupled
evolution equations, focussing on the history of the progenitor of
SDSS 1148+5251 and demonstrating the significant differences in the outcomes
for standard thin disk accretion models versus recent MHD models.

Our discussion is simplified and illustrative at best: much of the
input physics involving black hole seed formation, accretion flows, and mergers is still
being developed. Our main goal is to isolate some of the underlying 
local physical issues and parameters pertaining to accretion to better understand their role in
determining the global outcome of the cosmological Monte
Carlo simulations of SMBH build-up during hierarchical structure formation. 
Excellent overviews of the input physics have appeared elsewhere
(see, e.g. Haiman \& Quataert 2004, and references to earlier work).
Also, earlier treatments have considered some of the constraints on cosmological 
growth imposed by the recent discoveries of luminous quasars at high redshift
(e.g., Haiman \& Loeb 2001; Haiman 2004; Yoo \& Miraldi-Escud\'{e} 2004). 
But here we specifically want to illustrate
in the simplest fashion how the most recent findings related to relativistic, 
MHD accretion flows onto spinning black holes have important 
implications for evolutionary models of the growth of SMBHs 
in the early universe. We emphasize by concrete example the point 
made in Gammie et al. (2003)
that tracking the spin as well as the mass
of a black holes is necessary to determine its 
growth (see also Volonteri et al. 2004).  We also show that
the value of accretion radiation efficiency, $\sim 0.1$, adopted in many Monte Carlo simulations
may not be entirely consistent with the latest MHD accretion disk modeling.
Determining this parameter, on which the outcome of cosmological
simulations of SMBH growth depends very sensitively (exponentially!), is
coupled to the spin evolution of the hole; both may now be within our grasp
via detailed relativistic MHD simulations of BH accretion.

The calculations performed here
are prototypical only; our main aim is to motivate the incorporation of these parameters in 
more detailed Monte Carlo studies and thereby sharpen some of the rules that enter these
simulations. We also hope to provide those members of the
computational relativistic MHD community who may not be  
SMBH Monte Carlo cognoscenti
a simple means of extracting the essence of the
Monte Carlo simulations, particularly the 
evolutionary consequences of different accretion models.

In Section 2 we set out the basic equations that describe the
coupled evolution of black hole mass and spin by accretion. We also
summarize in this section the relations we require from the concordance $\Lambda$CDM cosmological
model. In Section 3 we integrate the coupled evolution equations to
track the evolution of black hole mass and spin as functions of time. 
In Section 4 we apply the results to the cosmological problem and
identify some constraints imposed by the existence of a SMBH at $z \geq 6.43$.
In Section 5 we summarize briefly and discuss some caveats and areas for
futher study.

\section{Basic Equations}
\label{Sec2}

Here we assemble the fundamental
black hole accretion evolution equations and
review the underlying
assumptions on which they are based. We then specify a
background cosmological model in the which the growth
of the black hole to supermassive size occurs.

\subsection{Black Hole Mass and Spin Evolution}
\label{accrete}

Define $\epsilon_M$, 
the efficiency of conversion of rest-mass energy
to luminous energy by accretion onto a black hole of mass $M$, according to
\begin{equation} \label{epsM}
\epsilon_M \equiv L/\dot M_0 c^2, 
\end{equation}
where $\dot M_0$ is the rate of rest-mass accretion and $L$ is the luminosity. Define 
$\epsilon_L$, the efficiency of accretion luminosity, 
according to 
\begin{equation} \label{epsL}
\epsilon_L \equiv L/L_E,
\end{equation}
where $L_E$ is the Eddington luminosity, given by
\begin{equation} \label{Edd}
L_E =\frac{4 \pi M \mu_e m_p c}{\sigma_T}
\approx 1.3 \times 10^{46} \mu_e M_8 ~{\rm erg~s^{-1}}.
\end{equation}
Here we assume that the accretion is dominated by normal, baryonic matter and
ignore any contribution of collisionless or self-interacting dark matter
(but see, e.g.,  Ostriker 2000 and  
Balberg \& Shapiro 2002 for alternative scenarios). 
In particular, we assume that the accreting gas consists of 
fully ionized atoms and that the principal
opacity source is Thomson scattering. The quantity $\mu_e$ is the mean
molecular weight per electron and $m_p$ is the proton mass.
The black hole growth rate must account for the loss of accretion mass-energy 
in the form of outgoing radiation according to
\begin{equation} \label{grwth}
\frac{dM}{dt} = (1-\epsilon_M) \dot M_0.
\end{equation}
Combining eqns.~(\ref{epsM})--(\ref{grwth}) we obtain the
black hole growth rate,
\begin{equation} \label{mdot}
\frac{dM}{dt} = \frac{\epsilon_L (1-\epsilon_M)}{\epsilon_M}\frac{M}{\tau},
\end{equation}
where $\tau$ is the characteristic accretion timescale,
\begin{equation} \label{tau}
\tau \equiv \frac{M c^2}{L_E} \approx 0.45\mu_e^{-1} ~{\rm Gyr},
\end{equation}
and is independent of $M$.
 
The mass accretion efficiency $\epsilon_M$ is typically a function of the black
hole spin parameter $a/M=J/M^2$. It changes with time as the spin evolves.
It is convenient to express the spin evolution in terms of the nondimensional 
quantity $s = s(a/M)$, defined by
\begin{equation} \label{s}
s \equiv \frac{d(a/M)}{dt} \frac{M}{\dot M_0}.
\end{equation}
Inserting eqns.~(\ref{grwth}) and (\ref{mdot}) into eqn.~(\ref{s}) yields
the evolution equation for the black hole spin,
\begin{equation} \label{adot}
\frac{d(a/M)}{dt} = \frac{\epsilon_L}{\epsilon_M} \frac{s}{\tau}.
\end{equation}
In general, eqns.~(\ref{mdot}) and ({\ref{adot}) must be integrated
simultaneously to determine the mass and spin evolution of the black hole.

Determining $\epsilon_M(a/M)$ and $s(a/M)$, which are needed to integrate
eqns.~(\ref{mdot}) and ({\ref{adot}),
requires a  
gas dynamical model for black hole accretion. We shall assume that the gas
has sufficient angular momentum to form a disk about the hole and consider
two different accretion disk models: 
(1) a standard, relativistic, Keplerian ``thin disk'' with
``no-torque boundary conditions'' at the innermost stable circular
orbit (ISCO) (Pringle \& Rees 1972; 
Novikov \& Thorne 1973; see Shapiro \& Teukolsky 1983 for review and 
references) and (2) a relativistic,  MHD accretion disk that accounts for
the presence of a frozen-in magnetic field in a
perfectly conducting plasma (De Villiers \& Hawley 2003; 
De Villiers, Hawley \& Krolick 2003; Gammie, Shapiro \& McKinney 2003;
McKinney \& Gammie 2004; and references therein). 
In the MHD model the magnetorotational instability 
(MRI; Balbus \& Hawley 1991) drives magnetic turbulence 
and provides the necessary torque to remove angular momentum from the
gas and drive the inflow. The MHD model is arguably the
most realistic model for disk accretion of magnetized plasma onto a black hole.
The standard thin disk model provides a simple, analytic, 
limiting case that is useful as a point
of comparison.

In a standard thin disk corotating with the black hole, the energy and angular momentum 
per unit rest mass 
accreted by a black hole are the energy and angular momentum of a
unit mass at the ISCO, immediately prior to its rapid plunge and capture
by the hole: 
\begin{eqnarray} \label{isco}
\tilde E_{\rm ISCO} &=& 
\frac{r_{\rm ms}^2-2M r_{\rm ms}+a \sqrt{M r_{\rm ms}}}
{r_{ms}(r_{ms}^2 - 3 M_h r_{ms} + 2 a \sqrt{M
r_{ms}})^{1/2} } 
\\
\tilde l_{\rm ISCO} &=&
\frac{\sqrt{M r_{\rm ms}} (r_{\rm ms}^2 - 2 a \sqrt{M
r_{\rm ms}} + a^2) }{r_{ms}(r_{ms}^2 - 3 M_h r_{ms} + 2 a \sqrt{M
r_{ms}})^{1/2} }, \nonumber
\end{eqnarray}
where the ISCO radius $r_{\rm ms}$ is given by
\begin{equation}
r_{\rm ms} = M\{ 3 + Z_2 - [(3 - Z_1)(3 + Z_1 + 2Z_2)]^{1/2} \},
\end{equation}
where
\begin{equation}
Z_1 \equiv 1 + \left
(1 - \frac{a^2}{M^2}
\right )^{1/3}
\left [
\left (1 + \frac{a}{M} \right )^{1/3} +
\left (1 - \frac{a}{M} \right )^{1/3}
\right ],
\end{equation}
and
\begin{equation} \label{Z2}
Z_2 \equiv \left (3 \frac{a^2}{M^2} + Z^2_1
\right )^{1/2}
\end{equation}
(see, e.g., Shapiro \& Teukolsky 1983, eqns 12.7.17-12.7.18 and 12.7.24).
The mass accretion efficiency and spin evolution parameters corresponding to
the thin disk model are then given by
\begin{equation} \label{eps}
\epsilon_M = 1-\tilde E_{\rm ISCO},
\end{equation}
\begin{equation} \label{thin}
s = \tilde l_{\rm ISCO} -2 \frac{a}{M} \tilde E_{\rm ISCO} 
~~~~({\rm standard \ thin \ disk}). 
\end{equation}

The MHD disk accretion model of Gammie, Shapiro \& McKinney (2004)
and McKinney \& Gammie (2004)  
is based on a fully relativistic, axisymmetric
simulation of a nonradiative, magnetized plasma onto a Kerr-Schild black hole
within the MHD approximation. The initial gas configuration is a torus with
an inner radius at $r/M =6$; in the absence of a magnetic field the torus
is constructed to be in equilibrium about the hole (Fishbone \& Moncrief 1976).
The torus in threaded with a poloidal magnetic field initially and
evolves with an adiabatic equation of state (EOS) with 
an adiabatic index $\Gamma = 4/3$ (to model a radiation-dominated, inner-disk EOS). The
source of viscosity is MHD turbulence driven by the MRI instability.
The simulations are performed for various black hole spin parameters
$a/M$, holding the value of the spin parameter fixed during the simulation.
The evolution proceeds for many 
dynamical timescales $M$, until a crude steady-state, with fluctuations,
is achieved. 

The results of the numerical simulations suggest that in steady-state 
the radiation
efficiency parameter $\epsilon_M$ as a function of $a/M$ 
is remarkably close to the function characterizing the standard thin disk 
(eqn.~\ref{eps}), even though there is no sharp transition in the surface density at or near 
the ISCO. However, 
the spin evolution parameter $s(a/M)$ is different and can be represented reasonably well 
by the least squares linear fit
\begin{equation} \label{MHD}
s = 3.14 - 3.30 \frac{a}{M}
~~~~({\rm MHD \ disk}), 
\end{equation}
(see Table 2 in McKinney \& Gammie 2004). 
The numerical simulations demonstrate that the above parameters
describing steady-state, MHD accretion-disk behavior are not particularly sensitive
to the initial conditions in the disk (e.g. the initial $B-$field). As 
McKinney \& Gammie (2004) discuss, the key results are also quite
comparable to those found by De Villiers et al. (2004),
who used a different numerical method and took the adiabatic index of the gas to
be $\Gamma = 5/3$ instead of $\Gamma = 4/3$.  We therefore model a relativistic 
MHD accretion disk by adopting
eqns~(\ref{eps}) and (\ref{MHD}) in our evolution equations. 

The $s$ vs. $a/M$ curve for the MHD disk is roughly parallel to, but
somewhat below, the curve for the standard disk (see Fig 4 in Gammie,
Shapiro \& McKinney 2004 for a comparison). In particular, the parameter $s$ never falls
to zero for a standard thin disk until the hole is maximally rotating at $a/M =1$, 
while for an MHD disk 
$s$ crosses zero at $a/M \approx 0.95$.
The crucial physical consequence is that steady accretion via
a standard thin disk always causes the central black hole to spin up
until it is maximally rotating (Bardeen 1970), while accretion via an
MHD disk drives the black hole to spin equilibrium at $a/M \approx 0.95$.
While the value of $a/M$ at equilibrium is not determined precisely 
and may depend on details of the flow
geometry (and the presence of radiative cooling), its departure from maximal 
is robust and is typically $\gtrsim 0.05$ below unity. 
The difference between $a/M=1$ and $a/M=0.95$ is quite substantial physically:
for example, the efficiency paramter is $\epsilon_M = 1-1/3^{1/2} = 0.42$ for
$a/M = 1$, while it is considerably smaller, 
$\epsilon_M = 0.19$, for $a/M = 0.95$.
As we will see below, this difference will have significant consequences for the
cosmological growth of a black hole from an initial seed. For comparison
we note that the radiative efficiency for disk accretion onto a nonrotating
Schwarzschild black hole ($a/M=0$) is $\epsilon_M = 1 - (8/9)^{1/2} = 0.057$.

If we assume that the supply of gas remains sufficiently copious in the vicinity
of the black hole, then the accretion luminosity
is likely to be Eddington-limited and nearly constant, with
$\epsilon_L \approx 1$. Accretion models characterized by
super-Eddington luminosities with
$\epsilon_L > 1$ are possible theoretically
(Ruszkowski \& Begeleman 2003). However,
the estimated Eddington ratios of the bolometric luminosities of the
broad-line quasars in
a Sloan Digital Sky Survey sample of 12,698 quasars in the redshift
interval $0.1 \leq z \leq 2.1$ support the value $\epsilon_L \approx 1$
as a physical upper limit (McLure \& Dunlop 2004). (Here the mass of the central
black hole is estimated from a virial relation, assuming that the gas motions
in the broad-line region of the quasars, measured by H$\beta$ and
MgII emission lines, are virialized.) The maximum allowed value of
$\epsilon_L$ establishes the maximum growth rate of a black hole from accretion
(see eqn.~\ref{mdot}) and, as a result, 
the most robust (least stringent) constraints on any gas accretion scenario for the growth of a
supermassive black hole from a smaller intial seed.
To establish these
constraints we therefore shall set $\epsilon_L = 1$ in many of our numerical estimates below.
                                                                                      
As a point of reference, and for later application, it is useful to integrate eqn.~(\ref{mdot})  
assuming that the mass and luminosity efficiencies both remain constant with
time, yielding
\begin{equation} \label{amp}
M(t)/M(t_i)= {\rm exp} \left[\frac{\epsilon_L (1-\epsilon_M)}{\epsilon_M}
\frac{(t-t_i)}{\tau} \right], ~~~~~(\epsilon_M, \epsilon_L \ {\rm constant}),
\end{equation}
where $t_i$ is the initial time at which the
black hole has a mass $M_i$. 
Note that the right hand side of eqn.~(\ref{amp}) is independent of black hole mass.
This fact makes it possible, under certain plausible conditions that we shall specify, 
to disentangle and track separately the amplification of black hole mass
by accretion from the amplification by discrete mergers.
Let $M_n(t)$ be the mass of the 
black hole at time $t$ following its $n^{th}$ merger with another hole at time $t_n$,
where $t_n \leq t \leq t_{n+1}$. Assume that the duration of a merger, as well as the
time required for accretion to drive the merged remnant 
to spin equilibrium (see eqn.~\ref{taugrw} below),
are both much shorter than the time interval between mergers, and that the hole continues to accrete
steadily throughout this interval. Note that black hole mergers can completely
eject black holes from halo centers owing to gravitational wave recoil, and thereby
turn off accretion altogether (see discussions of black hole recoil in halos in, e.g., 
Hut \& Rees 1992; Merritt et al. 2004; Madau \& Quataert 2004). However,
incorporating the
most recent recoil calculations (Favata, Hughes \& Holz 2004) 
into simple models of dark halo mergers, Yoo \& Miralda-Escud\'{e} (2004)
conclude that the kick velocities are not sufficiently large to impede black 
hole growth significantly (cf. Haiman 2004).
Note also that a major merger between two black holes of comparable 
mass may change 
both the magnitude and direction of the spin of the resulting black hole remnant. 
Following such a merger,
the orientation of the black hole spin may not be aligned 
with the orientation of the asymptotic gaseous disk at radii $r \gg 100M$ outside the hole. 
(Rees 1978 and Natarajan \& Pringle 1998 point out that the black hole exerts a torque on the asymptotic disk,
which also exerts a torque back on the hole, eventually forcing
their mutual alignment, but on a timescale that is still uncertain).
Moreover, the orientation of the asymptotic disk will likely fluctuate in time, due to the
redistribution of gas following mergers
of dark halo cores, galaxy mergers, and the tidal disruptions of passing stars by the central
hole. However, near the black hole,
at radii from $r \sim M$ to $r \sim 20M$, where the bulk of the disk's gravitational energy is
released and the hole-disk interactions are strong, the hole's gravitomagnetic field will exert a
force on the disk that, when combined with viscous forces and magnetic fields, 
will drive the disk down into the hole's equatorial plane,  
(the ``Bardeen-Petterson effect''; Bardeen \& Petterson 1975). This effect will maintain the 
alignment between the axis of the inner disk and the
spin axis of the hole. Accretion will subsequently drive the hole to spin equilibrium
and restore $\epsilon_M$ to its equilibrium value. 

For the situation described above we can take $\epsilon_M$ to
be constant and equal to its value at spin equilibrium throughout most of the lifetime
of the hole. Assume further that $\epsilon_L$ is also constant, which should be the case if the available
gas is sufficiently copious that the
accretion is always Eddington-limited, whereby $\epsilon_L \approx 1$. Let $f_n$ be the
mass amplification of the hole following its $n^{th}$ merger with another hole:
$f_n = M_n(t_n)/M_{n-1}(t_n) > 1$. Then we may use eqn.~(\ref{amp}) to calculate 
the total mass amplification from $t_i$ to $t_f$ according to
\begin{eqnarray} \label{merge}
\frac{M_f}{M_i} &=&
\frac{M_N(t_f)}{M_0(t_i)} =
\frac{M_0(t_1)}{M_0(t_i)} \ \frac{M_1(t_1)}{M_0(t_1)} \ \frac{M_1(t_2)}{M_1(t_1)} \cdots \\
& & \cdots \frac{M_{N-1}(t_N)}{M_{N-1}(t_{N-1})} \ \frac{M_N(t_N)}{M_{N-1}(t_N)} \ \frac{M_N(t_f)}{M_N(t_N)} 
\nonumber  \\
&=& {\rm exp} \left[C(t_1-t_i) \right]
\ f_1 \
 {\rm exp} \left[C(t_2-t_1) \right]
\cdots \\
& &  \cdots {\rm exp} \left[C(t_N-t_{N-1}) \right]
\ f_N \
{\rm exp} \left[C(t_f-t_N) \right]
\nonumber
\\
&=& f_1 f_2 \cdots f_N \ {\rm exp} \left[C(t_f-t_i) \right],
\nonumber
\end{eqnarray}
where $C=\epsilon_L (1-\epsilon_M)/(\epsilon_M \tau)$ and is essentially constant. 
Comparing  eqns.~(\ref{amp}) and (\ref{merge})
reveals that the net mass amplification due to accretion can be treated as a single 
multiplicative factor 
that is independent of the net amplification factor due to mergers, $f=f_1 f_2 \cdots f_N$.
Thus, for the simple scenario envisioned here, 
eqn.~(\ref{amp}) can be applied to determine the growth of a black hole by
accretion, even when steady growth by accretion 
is interrupted and augmented by discrete, stochastic black hole mergers.

\subsection{Cosmological Model}
\label{cosmo}

To relate the time parameter $t$ appearing in the evolution equations to
an observable time-like quantity like the redshift $z$,  
we adopt the concordance $\Lambda$CDM, spatially flat ($k=0$), 
cosmological model. All the free parameters in this model that we
shall need for our computations 
have been measured by now, so that the model is uniquely specified.

The basic evolution equation for the Friedmann-Robertson-Walker expansion parameter $a(t)$ 
is given by
\begin{equation} \label{FRW}
H^2 \equiv \left( \frac{\dot a}{a} \right)^2 = 
H_0^2 \left[ \Omega^0_m \left( \frac{a_0}{a} \right)^3 + 
     \Omega^0_\Lambda \right],
\end{equation}
where $H(t)$ is Hubble's constant and the normalized mass density parameter $\Omega^0_m$
and cosmological constant parameter $\Omega^0_\Lambda$ satisfy the relation
\begin{equation} \label{omega}
\Omega^0_m + \Omega^0_\Lambda = 1.
\end{equation}
In the above equations the sub(super)script ``0'' denotes the value of a quantity 
at the current epoch, $z=0$.
Recalling that $a/a_0 = 1+z$ and substituting eqn.~(\ref{omega}) into
eqn.~(\ref{FRW}), we can integrate eqn.~(\ref{FRW}) to obtain $t = t(z)$:
\begin{equation} \label{z}
t(z) = \frac{2}{3 H_0 (1-\Omega_m^0)^{1/2}}
{\rm sinh}^{-1}\left[\left (\frac {1-\Omega_m^0}{\Omega_m^0} \right)
\frac{1}{(1+z)^3} \right]^{1/2}.
\end{equation}
To evaluate $t(z)$ appearing above we adopt the WMAP values for the 
cosmological parameters: 
$\Omega^0_m \approx 0.27$ and $H_0 \equiv 100 h$ km/s/Mpc
with $h \approx 0.71$ (Bennett et al. 2003; Spergel et al. 2003).

To evaluate the molecular weight $\mu_e$ appearing in eqns.~(\ref{Edd}) and (\ref{tau})
we assume that the accreting plasma is a zero-metallicity, primoridial gas, for which
\begin{equation} \label{mue}
\mu_e = \frac{1}{1-Y/2}.
\end{equation}
We take the primordial helium abundance $Y$ 
to be $Y \approx 0.25$ (Cyburt, Fields \& Olive 2003).

For the concordance model, the age of the universe is $T=t(0)=13.7$ Gyr,
while the age at redshift $z=6.43$, the highest known quasar redshift,
(SDSS 1148+525; Fan et al. 2003), is only $t_{\rm QSO} = 0.87$ Gyr.
Hence $t_{\rm QSO}$ is the upper limit to the time available for accretion to occur onto the
initial seed black hole that powers this quasar. In fact, assuming that the black hole seed forms 
from the collapse of a first generation, Pop III star at
redshift $z \la 40$ and $t \ga t(40)= 0.067$ Gyr, the available time for accretion is reduced 
to $t_{\rm accrete} \la 0.80$ Gyr. 
We note that the stellar evolution (hydrogen burning) lifetime of a massive Pop III
progenitor, $t_{\rm evo} \sim 0.003$ Gyr (Wagoner 1969; Onukai \& Palla 2003),
is much smaller than $t(40)$, so that the delay between stellar formation and collapse is of little
consequence for determining the total time available for accretion growth.
Most important, the exponential accretion growth timescale is given by eqns.~(\ref{mdot}),
(\ref{tau}) and ({\ref{mue}) to be
\begin{equation} \label{taugrw}
\tau_{\rm grwth} = \frac{\epsilon_M}{1-\epsilon_M} \frac{1}{\epsilon_L} \tau =
0.0394 \frac{(\epsilon_M/0.1)}{1-\epsilon_M} \frac{1}{\epsilon_L} {\rm \ Gyr,}
\end{equation}
which is considerably smaller than $t_{\rm accrete}$.
It is the availability of many exponential growth timescales
from the time of black hole seed formation to the birth of a quasar that
makes it possible for the black hole to grow 
from stellar to supermassive size by gas accretion (aided by mergers) 
in the early universe.

\section{Black Hole Growth and Spinup}
\label{Sec3}

Here we integrate the coupled mass and spin evolution equations 
~(\ref{mdot}) and (\ref{adot})
to study black hole growth and spin-up by gas accretion. 
In Fig. 1 we show the increase in mass by accretion at the Eddington limit 
($\epsilon_L = 1$) as a function of time. 
In (a) the accretion disk is a standard thin disk with a 
``no torque boundary condition'' at the ISCO (eqn.~\ref{thin}); in (b) the
disk is an MHD disk as calculated by McKinney \& Gammie (2004) (eqn.~\ref{MHD}). 
We consider two different
initial values for the black hole spin parameter, $a/M$. The solid lines
show the case in which
the initial black hole is nonrotating with $a/M = 0$.
The dotted lines show the case in the which
the black hole to be spinning with $a/M = 0.75$. 
The latter
is the value calculated for a black hole formed from the catastrophic 
collapse of a
massive, radiation-dominated star 
spinning uniformly at the mass-shedding limit that has evolved to the onset
of relativistic radial instability just prior to collapse (Shibata \& Shapiro 2002; 
Shapiro \& Shibata 2002; Shapiro 2004).
For each disk model, the solid dots show the black hole growth
that would occur if the accretion were maintained at the asymptotic black hole spin and 
efficiency from the beginning.
Time is plotted in units of $\tau$ given by eqns.~(\ref{tau}) and (\ref{mue})
for a zero-metallicity, cosmological abundance of H and He.
The total duration plotted corresponds to the age difference 
between redshift $z_i = 40$ 
(the earliest plausible birthdate of a black hole from the collapse of a Pop III star) and
redshift $z_f = 6.43$ (the highest known quasar redshift, corresponding to SDSS 1148+5251; Fan et al. 2003) 
in the adopted $\Lambda$CDM cosmology.

\begin{figure}
\plotone{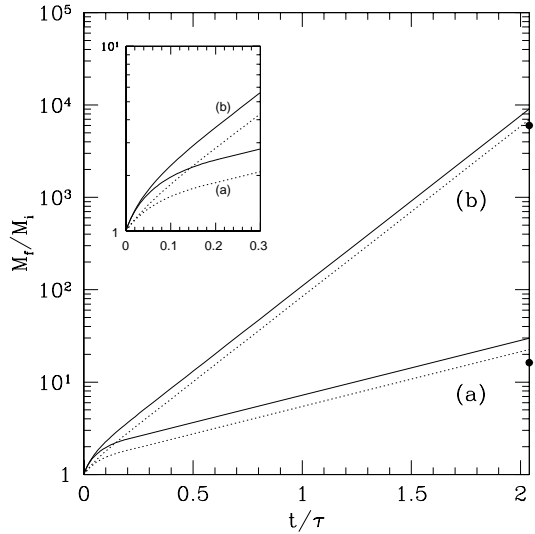}
\caption{ Black hole growth by accretion at the Eddington limit.
In (a) the accretion disk is a standard thin disk with a
``no torque boundary condition'' at the ISCO, ignoring photon recapture; in (b) the
disk is an MHD disk as calculated by McKinney \& Gammie (2004).
The solid lines show evolution starting from
$a/M = 0$, while the dotted lines show evolution starting from
$a/M = 0.75$.  For each disk model, the solid dots show the mass amplification
that would occur if the accretion were maintained at the asymptotic, equilibrium
spin and radiation efficiency from the beginning.
Time is plotted in units of $\tau = Mc^2/L_E = 0.45/ \mu_e$ Gyr. 
The total duration shown corresponds to the age difference between redshift $z_i = 40$
(the earliest plausible birthdate of a black hole from the collapse of a Pop III star) and
redshift $z_f = 6.43$ (the highest known quasar redshift, corresponding to 
1148+5251) in a concordance $\Lambda$CDM cosmology.
\label{Fig1}}
\end{figure}

The key difference between the two disk models is that the standard thin disk drives the black hole
to maximal spin and efficiency ($a/M = 1$ and $\epsilon_M = 0.42$; Bardeen 1970) while the MHD disk drives the
black hole to spin equilibrium at lower values ($a/M \approx 0.95$ and $\epsilon_M \approx 0.19$).
{\it A moderately lower radiation efficiency $\epsilon_M$ 
for Eddington-limited accretion results in substantially 
larger black hole growth.}  

As is evident from the figure, apart from an initial transient lasting 
$ \sim 0.1 \tau \sim \tau_{\rm grwth}$
(see figure inset), {\it the asymptotic evolution and final black hole mass amplification
factor is little affected by the initial black hole spin.}

In Fig. 2 we show the spin evolution during Eddington-limited accretion for
the cases shown in Fig. 1. 
The large figure plots the spin parameter versus time, while
the inset plots the spin parameter as a function of black hole mass amplification.
After the initial transient lasting $\sim 0.1 \tau$, during which time the
black hole grows by a factor of $\sim 2$, the black hole spin and 
efficiency approach their asymptotic values. An analytic integration 
of eqn.~(\ref{adot}) with eqn.~(\ref{MHD}) , crudely 
holding $\epsilon_M$ fixed, yields an exponential spin-up timescale 
$\tau_{\rm spin} \approx (1-\epsilon_M) \tau_{\rm grwth}/3.30$, which
explains the rapid spin-up rate. This result also explains why 
the asymptotic black hole evolution 
and mass amplification
are little affected by the initial black hole spin and depend
only on the equilibrium spin rate of the accreting Kerr black hole.
Thus, for all but the initial transient, it is adequate to adopt the equilibrium spin rate and 
use eqn.~(\ref{amp}) with  
the corresponding constant 
efficiency $\epsilon_M[(a/M)_{\rm eq}]$ to calculate the cosmological 
black hole mass amplification in lieu of a detailed integration of coupled 
evolution equations. 

\begin{figure}
\plotone{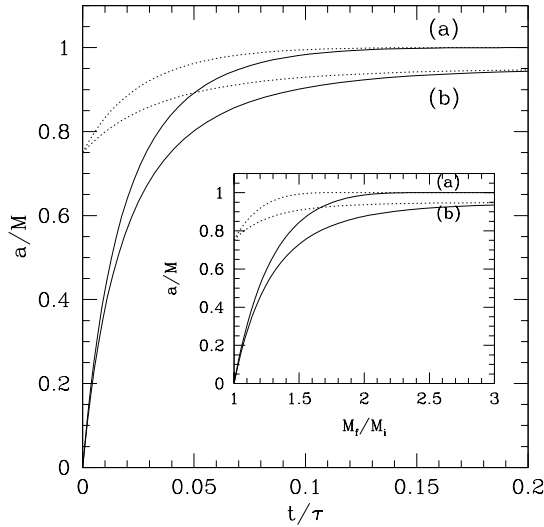}
\caption{
Spin evolution during Eddington-limited accretion for
the cases shown in Fig 1.  Curves are labelled as in that figure.
The large figure plots the spin parameter versus time, while
the inset plots the spin parameter versus the black hole mass amplification factor.
\label{Fig2}}
\end{figure}

\section{Cosmological Implications}
\label{impl}

Next we consider some of the cosmological implications of the results of
the previous section for the growth of SMBHs in the early 
universe.
 
In Fig. 3 we employ eqns.~(\ref{amp}) and (\ref{z}) to study the
black hole mass amplification factor $M_f/M_i$ as a function of the 
redshift $z_i$ of the initial seed black hole. Here we plot the 
final amplification factor achieved by
redshift $z_f = 6.43$, the highest known quasar redshift.
Each solid curve is labelled by the adopted
constant radiation
efficiency, $\epsilon_M$; the luminosity is assumed to be the Eddington
value ($\epsilon_L =1$). The horizontal dashed lines indicate the
range of amplification factors required for {\it accretion alone} to grow a
supermassive black hole of mass $10^9 M_{\odot}$ 
from an initial seed black hole of mass 
$100 \le M/M_{\odot} \le 600$ formed from the collapse of a Pop III star.
A mass of $10^9 M_{\odot}$ is the value inferred for typical quasars, including
1148+5251 (Fan et al. 2003).
The lower the mass of the initial seed, the larger is the required mass
amplification. 
The horizontal dotted lines indicate the required range of accretion-driven 
mass amplification assuming that {\it mergers} assist the 
growth process by accounting
for an amplification of $f \sim 10^4$ in black hole mass, with
the remaining amplication provided by gas accretion 
(Yoo \& Miralda-Escud\'{e} 2004).

\begin{figure}
\plotone{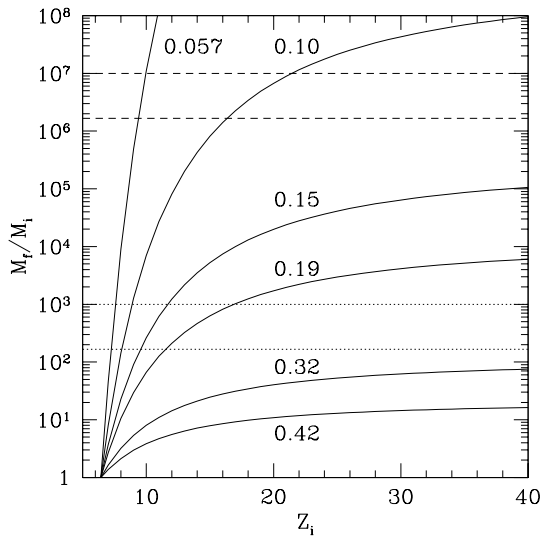}
\caption{
Black hole accretion mass amplification $M_f/M_i$ versus redshift $z_i$
of the initial seed. Here we plot the amplification achieved by
redshift $z_f = 6.43$, the highest known quasar redshift, corresponding to
1148+5251. Each solid curve is labelled by the adopted
constant radiation efficiency, $\epsilon_M$; the luminosity is assumed to be the Eddington
value ($\epsilon_L =1$). The horizontal dashed lines bracket the
range of amplification required for accretion alone to grow a
seed black hole of mass
$100 \le M/M_{\odot} \le 600$ formed from the collapse of a Pop III star
to $10^9 M_{\odot}$.
The horizontal dotted lines bracket the required accretion amplification range  
assuming that mergers account for a growth of $10^4$ in black hole mass,
the remainder being by gas accretion.
\label{Fig3}}
\end{figure}

Some important cosmological implications can be inferred from 
Fig 3.  Consider a seed black hole that forms sometime after
redshift $z_i \la 40$, by which
time the earliest stars have formed and collapsed.
In the absence of mergers, steady accretion cannot by itself 
achieve the required growth to explain quasars at $z_f = 6.43$ 
unless the efficiency satisfies $\epsilon_M \la 0.13$.
For disk accretion, 
this constraint requires the steady-state black hole spin parameter to be 
below $a/M \approx 0.83$. This value is below (1) the maximal
black hole spin $a/M = 1$, the asymptotic equilibrium value for a 
standard thin disk; (2) $a/M = 0.998$ (for which $\epsilon_M \approx 0.32$), 
the equilibrium value of a standard 
thin disk accounting for the recapture of some of the emitted photons by the
black hole (Thorne 1974); and (3) $a/M \approx 0.95$, the equilibrium value of a
typical MHD disk. Therefore, it is likely that {\it mergers are required to
assist accretion to achieve black hole growth to supermassive size 
by $z_f = 6.43$}.  As may be inferred from the figure, this conclusion holds 
even in the (unlikely) event that
the black hole seed forms much earlier
than $z_i \approx 40$. 

Monte Carlo simulations by 
Yoo \& Miralda-Escud\'{e} (2004)
of hierarchical CDM halo mergers, accompanied by
mergers of their central black holes, suggest that black hole mass amplification
factors of $\sim 10^4$ are achieved via mergers by $z_f = 6.43$.
This result implies that 
$f = f_1 f_2 \cdots f_N \sim 10^4$ in eqn.~(\ref{merge}).
While other simulations predict lower growth factors due to mergers, we adopt
$10^4$ to be conservative, noting that our conclusions below are strengthened if the smaller
factors turn out to be correct. 
As eqn.~(\ref{merge}) shows,
eqn.~(\ref{amp}) 
can be used to estimate
the cumulative mass amplication by gas accretion, 
even when steady growth by accretion 
is interrupted and enhanced by stochastic black hole mergers.
Minor mergers of massive black holes with smaller holes tend to
spin-down the massive hole (Hughes \& Blandford 2003; Gammie, 
Shapiro \& McKinney 2004). Major mergers of two black holes 
of comparable mass following binary inspiral 
drive the the merged remnant to $a/m \ga 0.7$. [See, e.g., 
 Baumgarte \& Shapiro 2003 for a review of binary black hole coalescence, and references; 
see Gammie, Shapiro \& McKinney 2004 for a general discussion of black hole 
spin-up and spin-down mechanisms, and references]. In either case, after
a short transient epoch, accretion will drive the merged remnant 
to the disk accretion equilibrium
spin rate and corresponding mass efficiency (see Section 2), as assumed
in employing eqn.~(\ref{amp}).

Fig 3. indicates that for black hole seeds that arise from collapsed
Pop III stars, merger-assisted mass amplification 
to SMBH status is easily achieved for 
typical MHD accretion disks, marginally possible for a standard thin
disk in spin equilibrium accounting for photon recapture, 
but not possible for a standard
thin disk that drives the black hole to maximal rotation.
However, {\it if the black hole seed is less than $600 M_{\odot}$, the standard thin
disk appears to be ruled out}: the required amplification places 
the lower dotted line in the plot above the curve 
for $\epsilon_M = 0.32$, corresponding to $a/M = 0.998$, the value for
a standard thin disk in spin equilibrium, accounting for photon recapture.
These conclusions again also hold
even if the
black seed forms much earlier
than $z_i \approx 40$, while they may be tightened if the seed forms later.
In fact, it may be likely that the seed forms later, at $z_i \la 40$, given
that even $4-\sigma$ peaks in the density perturbation spectrum
for the progenitor halo of SDSS 1148+5251
do not collapse until $z \sim 30$ in the $\Lambda$CDM concordance cosmology
(see, e.g. Figure 5 in Barkana \& Loeb 2001). Moreover, the potential wells of
the earliest halos are quite shallow ($\sim 1$ km/s) and may not be able to retain
enough gas to form stars. Nevertheless, the effect of altering the date of birth
of the BH seed is not very great unless $z_i \la 20 - 25$, as is evident from
Fig. 3. 

It is significant that {\it the range of equilibrium accretion disk radiation 
efficiences required 
to achieve the necessary 
growth of a black hole seed to supermassive size by $z_f = 6.43$ is consistent with the
values inferred observationally 
for $R$}, the ratio of the QSO plus AGN luminosity density to the mass density of 
SMBHs in nearby galaxies:
$\epsilon_M \ge R \sim 0.1 - 0.2$ (Soltan 1982;
Yu \& Tremaine 2002; Elvis et al. 2002).  
This consistency supports the notion that accretion
plays an important role in SMBH growth and is responsible for the 
acquisition of the 
bulk of the final mass of a SMBH. 
{\it The range of $\epsilon_M$ inferred from R favors
accretion disk models that drive the black hole to spin equilibrium in the
range $0.7 \la a/M \la 0.95$, well below maximal rotation and 
consistent with the values 
found in recent simulations of relativistic MHD accretion disks}.

How are our conclusions altered if (when) a quasar is discovered at
a higher redshift, $z_{\rm QSO} > 6.43$? We anticipate this possibility in
Fig.\ 4, where we solve eqn.~(\ref{amp}) to plot the mass 
efficiency $\epsilon_M$ required to build
a supermassive black hole from a seed as a function of  
the host quasar redshift, $z_{\rm QSO}$. Here we again assume that a seed black hole 
of mass $100 \le M/M_{\odot} \le 600$, formed from the collapse of a 
Pop III star at $z_i \gg z_{\rm QSO}$, 
grows by $z_{\rm QSO}$
to $10^9 M_{\odot}$.
While accreting, 
the black hole again is assumed to radiate at the
Eddington luminosity, with $\epsilon_L =1$.  
We set $t_i = 0$, $z_i = \infty$, so that the plotted radiation 
efficiency represents an
{\it upper} limit to $\epsilon_M$; increasing $t_i$  and lowering $z_i$ 
decreases the required 
efficiency (but the decrease is small if, for example, we set $z_i \ga 40$).
The dashed lines
indicate the range of efficiencies required for gas accretion alone to
achieve the necessary growth. The dotted lines indicate the range
required for gas accretion assuming that mergers assist the growth and 
account for an amplification of $f=10^4$ in black hole mass. 
The horizontal solid lines 
bracket the range of efficiencies inferred observationally for $R$.

\begin{figure}
\plotone{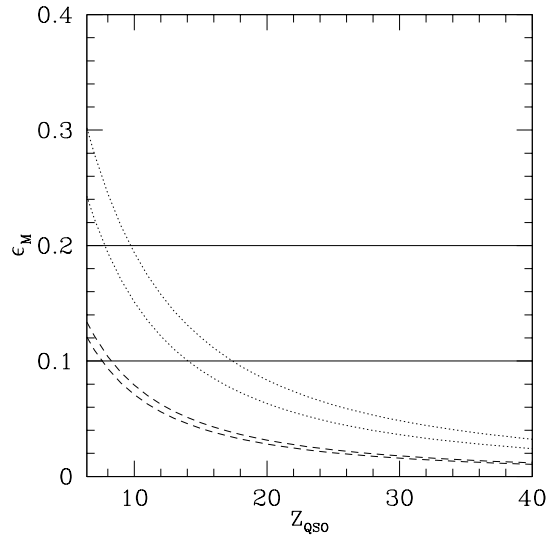}
\caption{
The radiation efficiency $\epsilon_M$ required to build a SMBH
versus the redshift $z_{\rm QSO}$ of the host quasar.
The seed black hole forms with a  mass $100 \le M/M_{\odot} \le 600$ 
from the collapse of a Pop III star at $z_i \gg z_{\rm QSO}$
and grows by $z_{\rm QSO}$ to $10^9 M_{\odot}$. Accretion occurs at the 
Eddington luminosity.  The dashed lines
bracket the range of efficiencies required for accretion alone to
achieve the necessary growth. The dotted lines bracket the range
required for accretion assuming that mergers account for a
growth of $10^4$ in black hole mass.
The horizontal solid lines
bracket the range of efficiencies inferred from $R$, the observed ratio of
the QSO plus AGN luminosity density to the mass density of
local SMBHs.
\label{Fig4}}
\end{figure}

In the absence of mergers, the upper limit to the 
radiation efficiency required to build a supermassive
black hole by $z_{\rm QSO} \approx 6.43$ is $\epsilon_M \approx 0.14$, 
consistent with observational
constraints implied by $R$, but inconsistent with the much higher values 
expected for a standard thin
disk driven to spin equilibrium, and marginally inconsistent for 
an MHD accretion disk. The inconsistencies become worse as 
$z_{\rm QSO} > 6.43$. In fact, the upper limit to $\epsilon_M$ falls below
0.1 for $z_{\rm QSO} \ga 8$, which would force $\epsilon_M$ to lie 
below the observationally inferred range for $R$ and would therefore be 
difficult to understand.

When mergers are included, the upper limit to the 
efficiency required to build a
SMBH by $z_{\rm QSO} \approx 6.43$ increases
to $\epsilon_M \sim 0.30$ (i.e., any black hole seed born at a finite redshift,
$\infty > z_i > z_{\rm QSO}$, must accrete at a lower efficiency
than $0.30$ to reach $10^9 M_{\odot}$).
This upper limit is roughly consistent with the observational
constraint on $\epsilon_M \geq R$ 
and with theoretical values for 
accretion from an MHD disk in spin equilibrium, but
only marginally consistent for accretion from 
a standard thin disk in spin 
equilibrium accounting for photon recapture,
and inconsistent for accretion from 
a standard thin disk that drives
the black hole to maximal rotation. {\it Should a quasar be discovered at 
$z_{\rm QSO} > 6.43$}, it would appear that 
{\it accretion from a standard thin disk will be ruled out}: the upper limit
will fall below $\epsilon_M = 0.32$, the value for
a standard thin disk in spin equilibrium accounting for photon
recapture.
{\it Should a quasar be discovered at
$z_{\rm QSO} \ga 10$}, $\epsilon_M$ would fall below $0.19$ and 
{\it the results would be difficult to reconcile 
with accretion from a typical MHD disk} as modeled in recent simulations. 
These critical values of
$z_{\rm QSO}$ are all smaller if the initial black hole seed
is smaller than $600 M{_{\odot}}$, as the top curve, which sets the limit,
is lowered in the figure.
Finally, should a quasar be discoverd (perhaps unexpectedly) 
at $z_{\rm QSO} \ga 18$, the upper limit to $\epsilon_M$ 
would drop below the observationally
inferred value $0.1$ for all $z_i \ge z_{\rm QSO}$. This drop would not 
be easily explained by any of these models of accretion, and might require
super-Eddington accretion with $\epsilon_L > 1$ to keep $\epsilon_M \gtrsim 0.1$.
Alternatively, the value of $R$, observed for local host galaxies, might
be significantly smaller at high redshift, which would relax the constraint
on $\epsilon_M$. But achieving the inferred lower radiation efficiencies
might then require ADAF disks or spherical accretion.

\section{Summary}
\label{sum}

We have explored the evolution of
black hole mass and spin by gas accretion in the early universe.
We have illustrated how for Eddington-limited accretion, the growth of a SMBH 
depends sensitively on the radiative efficiency, $\epsilon_M$. For disk accretion,
the mean efficiency is determined by the equilibrium black hole spin, which in turn depends on the
the torques acting on the gas near the black hole horizon. We have explored 
the consequences of the assumptions that seed black holes are the remnants
of collapsed Pop III stars that may form as early as $z \la 40$ and can 
grow to $\sim 10^9 M_{\odot}$ by $z_{\rm QSO} = 6.43$, the highest redshift 
discoverd to date, corresponding to QSO SDS 1148+5251. Allowing for growth both 
by accretion and mergers, simple theory suggests that the required mass amplification
is possible provided the radiation efficiency satisfies 
$\epsilon_M \la 0.2$, with the upper limit decreasing
should a quasar be discoverd at higher redshift $z_{\rm QSO} > 6.43$.
The inferred efficiency is consistent with the observed ratio $R$ of the QSO plus AGN radiation
density to the mass density of SMBHs in local galaxies, which 
suggests that an appreciable fraction of the mass of the final black hole is acquired
by disk accretion, rather than by mergers. The inferred efficiency favors MHD 
accretion disk models that
exert non-zero torques on the gas at the inner edge of the disk. These disks
ultimately drive the black hole to spin equilibrium at $a/M \sim 0.95$, 
substantially below the maximum spin allowed for a Kerr black hole.

We have made a number of simplying assumptions in an effort to provide a concrete
calculation that illustrates how the SMBH initial seed, its cosmological growth, and accretion 
disk models may all be constrained by the existence of quasars at high redshift. 
Our conclusions are tentative, as there exist many caveats. For example, if the progenitors of 
black hole seeds are
SMSs with $M \gg 10^3 M_{\odot}$ instead of Pop III stars, many of the constraints on the
radiation efficiency
$\epsilon_M$, and the associated range of expected black hole spins, may have to be
relaxed.  On the other hand, if the progenitors have masses $\la 100 M_{\odot}$ the constraints 
favoring low efficiencies $\epsilon_M \la 0.2$ characterizing MHD accretion disks in
spin equilibrium are strengthened. If the accretion luminosity is super-Eddington with $\epsilon_L > 1$, then the
upper limit for $\epsilon_M$ increases (see eqn.~\ref{amp}), relaxing the constraints which
favor an MHD disk over a standard thin disk. Super-Eddington accretion is
possible theoretically (Ruszkowski \& Begelman 2003). 
However, the a Sloan Digital Sky Survey of 12,698 broad-line quasars in the redshift
interval $0.1 \leq z \leq 2.1$ supports the Eddington value
as a physical limit (McLure \& Dunlop 2004).
Moreover, given the observed value for $R$,
a higher limit for $\epsilon_M$ would imply that a large fraction of the
emitted radiation lies outside the optical bandwidths included in the determination of $R$, which seems
unlikely. Finally, mergers may lead to black hole mass amplification factors smaller than 
$f \sim 10^4$, the typical value found in the Monte-Carlo 
calculations of Yoo \& Miralde-Escud\'{e}, and the value assumed here. 
A smaller merger amplification factor requires larger accretion-driven amplification
in order for a black hole seed to reach SMBH size by $z_{\rm QSO} = 6.43$ (see
eqn.~\ref{merge}). This would strengthen the constraints set by SMBH accretion growth on
the upper limit to $\epsilon_M$, although the efficiency must still be consistent with $R$ if
the bolometric accretion luminosity resides mostly in the optical wavebands included in $R$ and
the value of $R$ measured for local galaxies also applies at higher redshift. Of course,
the peak contribution to the value of $R$ arises from the average behavior of accreting 
BHs with $ z \sim 3$; this ratio 
does not necessarily apply to individual objects like SDSS 1148+5251, nor need
it apply to SMBHs at $z \approx 6$.

We have assumed that the mass of SDSS 1148+5251 is $\sim 10^9 M_{\odot}$; a lower value would relax
many of our constraints, which a higher value would strengthen them. If the flux from this 
source were amplified by gravitational lensing, or beaming, then a lower mass estimate would be appropriate. However, no multiple images have been seen (Richards et al. 2004) and it has been shown
that high amplification without at least two images is very improbable (Keeton et al. 2004).
Strong beaming also seems unlikely since it would reduce the line/continuum ratio 
(Haiman \& Cen 2002), which is not observed (Willott et al. 2003). In fact, assuming that
the quasar emits at the Eddington luminosity gives a mass of $4.6 \times 10^9 M_{\odot}$
(Fan et al. 2003; Haiman 2004), and this higher value strengthens our conclusions somewhat.

Isolating the accretion growth of a seed black hole in the early universe from the hole's full
dynamical history and environment does not allow us to account for other important
correlations that provide clues to the formation of SMBHs. Such correlations
include the SMBH  mass vs. bulge luminosity relation, $M_{\rm BH} \propto L_{\rm bulge}$ (Kormendy \& Richstone
1995), and the SMBH mass vs. velocity dispersion relation, 
$M_{\rm BH} \propto \upsilon_c^4$ (Gebhardt et al. 2000; Ferrarese \&
Merritt 2000; Tremaine et al. 2002), inferred for nearby host galaxies. Only by performing detailed
simulations that track the formation and growth of SMBHs in a cosmological 
setting governed by hierarchical halo mergers, black hole mergers, gas settling, star formation,
and feedback can these correlations be reliably reproduced. We look forward to
the next generation of simulations that incorporate the recent results of relativistic MHD accretion onto
black holes, since, as demonstrated here, the outcome of these global simulations 
may depend sensitively on the local physics of such accretion flows.

\acknowledgments

It is a pleasure to thank T. Abel, C. Gammie, J. McKinney, J. Miraldi-Escud\'{e}, 
and Q. Yu for valuable discussions. We also thank the referee for a critical reading of the
manuscript and valuable comments. A portion of this work was
performed during the summer of 2004 
at the Aspen Center for Physics, whose hospitality is 
gratefully acknowledged.  
This work was supported in part by NSF Grants 
PHY-0205155 and PHY-0345151 and NASA Grant NNG04GK54G at the
University of Illinois at Urbana-Champaign.

\end{document}